\documentclass[aps,pra,amsmath,amssymb,twocolumn,superscriptaddress,showpacs,nofootinbib]{revtex4}

\usepackage[bookmarks = true, pdfpagemode = None, pdfstartview = FitH, colorlinks = true, urlcolor = blue,hyperfootnotes=false]{hyperref}
\usepackage{graphicx}
\usepackage{color}

\usepackage[sort&compress]{natbib}
\newcommand{\leftexp}[2]{{\vphantom{#2}}^{#1}\!{#2}}
\newcommand{\CX}{\leftexp{C}{X}}
\newcommand{\CZ}{\leftexp{C}{Z}}
\newcommand{\C}[1]{\leftexp{C}{#1}}

\newcommand{\set}[1]{\mathcal{#1}}

\newcommand{\ket}[1]{| #1 \rangle}

\newcommand{\hhbar}{\mathcal{H}_l\cap\overline{\mathcal{H}}_r}
\newcommand{\hbarh}{\overline{\mathcal{H}}_l\cap\mathcal{H}_r}

\begin{document}

\title{Graphical description of the action of Clifford operators on stabilizer states}

\author{Matthew B.~Elliott}
\email{mabellio@unm.edu}
\affiliation{Department of Physics and Astronomy, University of New Mexico,
Albuquerque, NM 87131}

\author{Bryan Eastin}
\affiliation{Department of Physics and Astronomy, University of New Mexico,
Albuquerque, NM 87131}

\author{Carlton M.~Caves}
\affiliation{Department of Physics and Astronomy, University of New Mexico,
Albuquerque, NM 87131}

\begin{abstract}

We introduce a graphical representation of stabilizer states and
translate the action of Clifford operators on stabilizer states into
graph operations on the corresponding stabilizer-state graphs.  Our
stabilizer graphs are constructed of solid and hollow nodes, with
(undirected) edges between nodes and with loops and signs attached to
individual nodes.  We find that local Clifford transformations are
completely described in terms of local complementation on nodes and
along edges, loop complementation, and change of node type or sign.
Additionally, we show that a small set of equivalence rules generates
all graphs corresponding to a given stabilizer state; we do this by
constructing an efficient procedure for testing the equality of any
two stabilizer graphs.

\end{abstract}

\pacs{03.67.-a}

\maketitle

\section{Introduction}

Stabilizer states are ubiquitous elements of quantum information
theory, as a consequence both of their power and of their relative
simplicity.  The fields of quantum error correction,
measurement-based quantum computation, and entanglement
classification all make substantial use of stabilizer states and
their transformations under Clifford
operations~\cite{gottesmanthesis,oneway,vandennest71}.  Stabilizer states
are distinctly quantum mechanical in that they can possess arbitrary
amounts of entanglement, but the existence of a compact description
that can be updated efficiently sets them apart from other highly
entangled states. Their prominence, as well as their name, derives
from this description, a formalism in which a state is identified by
a set of Pauli operators generating the subgroup of the Pauli group
that stabilizes it, i.e., the subgroup of which the state is the
$+1$ eigenvector. In this paper we seek to augment the stabilizer
formalism by developing a graphical representation both of the states
themselves and of the transformations induced on them by Clifford
operations.  It is our hope that this representation will contribute
to the understanding of this important class of states and to the
ability to manipulate them efficiently.

The notion of representing states graphically is not new.  Simple
graphs are regularly used to represent \textit{graph states}, i.e.,
states that can be constructed by applying a sequence of
controlled-$Z$ gates to qubits each initially prepared in the state
$(\ket{0}+\ket{1})/\sqrt2$.  The transformations of graph states
under local Clifford operations were studied by Van den
Nest~\cite{vandennest69}, who found that local complementation
generated all graphs corresponding to graph states related by local
Clifford operations.  The results presented here constitute an
extension of work by Van den Nest and others to arbitrary stabilizer
states.

Our graphical depiction of stabilizer states is motivated by the
equivalence of stabilizer states to graph states under local Clifford
operations~\cite{vandennest69}. Because of this equivalence,
\textit{stabilizer-state graphs\/} can be constructed by first
drawing the graph for a locally equivalent graph state and then
adding decorations, which correspond to local Clifford operations, to
the nodes of the graph.  Only three kinds of decoration are needed
since it is possible to convert an arbitrary stabilizer state to some
graph state by applying one of six local Clifford operations
(including no operation) to each qubit.  The standard form of the
generator matrix for stabilizer states plays a crucial role in the
development of this material, particularly in exploring the
properties of \textit{reduced graphs}, a subset of stabilizer graphs
(which we introduce) that is sufficient for representing any
stabilizer state.  More generally, however, our stabilizer-state
graphs are best understood in terms of a canonical circuit for
creating the stabilizer state.  This description also permits the use
of circuit identities in proving various useful equalities.  In this
way, we establish a correspondence between Clifford operations on
stabilizer states and graph operations on the corresponding
stabilizer-state graphs.  Ultimately, these rules allow us to
simplify testing the equivalence of two stabilizer graphs to the
point that the test becomes provably trivial.

This paper is organized as follows.  Section~\ref{sec:background}
contains background information on stabilizer states, Clifford
operations, and quantum circuits.  Stabilizer-state graphs are
developed in Sec.~\ref{sec:graphs}, and a graphical description of
the action of local Clifford operations on these graphs is given in
Sec.~\ref{sec:transformations}.  The issue of the uniqueness of
stabilizer graphs is taken up in Sec.~\ref{sec:equiv}.  The appendix
deals with the graph transformations associated with $\CZ$ gates.

\section{Background}
\label{sec:background}
\subsection{Stabilizer formalism\label{subsec:stabilizer}}

The \textit{Pauli group\/} on $N$ qubits, $\mathcal{P}_N$, is defined
to be the group, under matrix multiplication, of all $N$-fold tensor
products of the identity, $I$, and the Pauli matrices, $X=\sigma_1$,
$Y=\sigma_2$, and $Z=\sigma_3$, including overall phases $\pm 1$ and
$\pm i$.  A \textit{stabilizer state\/} is defined to be the
simultaneous $+1$ eigenstate of a set of $N$ commuting, Hermitian
Pauli-group elements that are independent in the sense that none of
them can be written as a product of the others.  These elements are
called \textit{stabilizer generators\/} and are denoted here by
$g_j$, while $g_{jk}$ is used to denote the $k$th Pauli matrix in the
tensor-product decomposition of generator $g_j$.  Stabilizer
generator sets are not unique; replacing any generator with the
product of itself and another generator yields an equivalent
generating set.  An arbitrary product of stabilizer generators,
$g=g_1^{a_1}\cdots g_N^{a_N}$, where $a_j \in \{0,1\}$ is called a
\textit{stabilizer element\/}; the stabilizer elements make up a
subgroup of the Pauli group known as the \textit{stabilizer}.

A \textit{graph state\/} is a special kind of stabilizer state whose
generators can be written in terms of a simple graph as
\begin{equation}
\label{eq:graphgenerators}
g_j = X_j \prod_{k \in \set{N}(j)} Z_k\;,
\end{equation}
where $\set{N}(j)$ denotes the set of neighbors of node $j$ in the
graph (see Sec.~\ref{subsec:terminology} and Ref.~\cite{diestel} for
graph terminology). \textit{Simple graphs\/} and, hence, graph states
can also be defined in terms of an \textit{adjacency matrix\/}
$\Gamma$, where $\Gamma_{\!jk}=1$ if $j \in \set{N}(k)$ and
$\Gamma_{\!jk}=0$ otherwise.  In a simple graph, a node is never its
own neighbor, i.e., there are no self-loops; thus the diagonal
elements of the adjacency matrix of a simple graph are all equal to
zero.

\subsection{Binary representation of the Pauli group \label{subsec:binary}}

The binary representation of the Pauli group associates a
two-dimensional binary vector $r(\sigma_j)$ with each Pauli matrix
$\sigma_j$, where $r(I) = \left( \begin{array}{cc} 0 & 0
\end{array} \right)$, $r(X) = \left( \begin{array}{cc} 1 & 0
\end{array} \right)$, $r(Y) = \left( \begin{array}{cc} 1 & 1
\end{array} \right)$, and $r(Z) = \left( \begin{array}{cc} 0 & 1
\end{array} \right)$.
This association is generalized to an arbitrary element
$p\in\mathcal{P}_N$, whose $k$th Pauli matrix is $p_k$, by letting
$r(p)$ be a $2N$-dimensional vector whose $k$th and $(N+k)$th entries
are the entries of $r(p_{k})$, i.e., $r(p_{k})=\left(
\begin{array}{cc} [r(p)]_k & [r(p)]_{N+k}
\end{array} \right)$.  The binary representation of a Pauli-group
element specifies the element up to the overall phase of $\pm1$ or $\pm
i$; Hermitian Pauli-group elements are specified up to a sign.

The binary representation of the product of two Pauli-group elements,
$p,q\in\mathcal{P}_N$, is the binary sum of their associated vectors,
i.e., $r(p q) = r(p) + r(q)$.  Two such elements commute if
their \textit{skew product},
\begin{equation}
r(p)\wedge r(q)=\sum_{j=1}^N[r(p)]_j[r(q)]_{N+j}+[r(p)]_{N+j}[r(q)]_j\;,
\end{equation}
has value $0$; otherwise they anticommute.

Using binary notation, a full set of generators for a stabilizer
state can be represented (up to a sign for each generator) by an $N
\times 2N$ \textit{generator matrix\/} whose $j$th row is $r(g_j)$.
Because the stabilizer generators commute, the rows of the generator
matrix are orthogonal under the skew product.  Similarly, the
independence of the stabilizer generators under matrix multiplication
implies that the rows of the generator matrix are linearly
independent under addition.  The freedom to take products of
stabilizer generators without changing the stabilized state becomes,
for the generator matrix, the freedom to add one row of the matrix to
another.  The exchange of any pair of qubits $j$ and $k$ of a
stabilizer state corresponds to the exchange of columns $j$ and $k$
and columns $N+j$ and $N+k$ in the generator matrix. Since rows of
the generator matrix are linearly independent and have vanishing skew
product, these two operations are sufficient to allow us to transform
any generator matrix to a \textit{canonical
form}~\cite{nielsenchuang},
\begin{equation}
\label{eq:canonicalstabilizer}
\left(\begin{array}{cc|cc}I & A & B & 0 \\ 0 & 0 & A^T & I \end{array}\right)\;,
\end{equation}
where $B=B^T$ and the vertical line divides the matrix in half.  The
vanishing skew product of rows of the generator matrix implies that
$B$ is a symmetric matrix and that $A$ and $A^T$ appear as indicated.
The $I$s in Eq.~(\ref{eq:canonicalstabilizer}) denote a pair of
identity matrices whose dimensions sum to $N$. The dimension of the
upper left identity matrix is called the \textit{left rank\/} of the
generator matrix.  Due to the freedom inherent in qubit exchange, the
canonical form of the generator matrix is not unique.

For graph states, Eq.~(\ref{eq:canonicalstabilizer}) becomes
\begin{equation}
\label{eq:canonicalgraph}
\left(\begin{array}{c|c}I & B \end{array}\right)\;,
\end{equation}
where $B$ has only $0$s on the diagonal.  Graph states thus have
generator matrices of full left rank, with $B$ being the adjacency
matrix of the graph state's underlying graph.  We denote generator
matrices of this sort by the term \textit{strict graph form}, whereas
the term \textit{graph form} is used for generator matrices of the
form shown in Eq.~(\ref{eq:canonicalgraph}) where $B$ is any
symmetric matrix.

The binary representation does not encode the sign of Pauli group
elements, so the generator matrix really specifies a set of
generators up to $2^N$ possible sign assignments and thus specifies
not a single stabilizer state, but rather an orthonormal basis of
simultaneous eigenstates of the generators, each member of which
corresponds to one of the sign choices.  Despite this, we continue,
for convenience, to refer to \textit{the\/} stabilizer state
associated with a generator matrix.

\subsection{Clifford operations\label{subsec:clifford}}

The \textit{Clifford group\/} is the normalizer of the Pauli group,
i.e., the set of all unitary operators $U$ such that $UpU^{\dag} \in
\mathcal{P}_N$ for all $p \in \mathcal{P}_N$. Any local operation in
the Clifford group can be obtained by repeated application of
Hadamard and phase gates, which are written in the standard basis as
\begin{equation}
\label{eq:HSdefinition}
\begin{array}{ccc}
\displaystyle{H = \frac{1}{\sqrt{2}} \left(\begin{array}{cc}1 & 1 \\1 & -1\end{array}\right)}
&
\mbox{and}
&
\displaystyle{S= \left(\begin{array}{cc}1 & 0 \\0 & i\end{array}\right)\;,}
\end{array}
\end{equation}
respectively.  Adding the two-qubit controlled-$Z$ (or
controlled-sign) gate,
\begin{equation}
\label{eq:CZdefinition}
\begin{array}{c}
\CZ = \left(\begin{array}{cccc}1 & 0 & 0 & 0 \\0 & 1 & 0 & 0 \\0 & 0 & 1 & 0 \\0 & 0 & 0 & -1\end{array}\right)\;,
\end{array}
\end{equation}
completes the basis for Clifford operations~\cite{gottesman,aaronson}.

\subsection{Quantum circuits\label{subsec:circuit}}

Quantum-circuit notation~\cite{nielsenchuang} is a pictorial method
for representing the application of discrete operations to a quantum
system.  As with electrical circuits, repeated usage of a small
number of simple, standard parts results in complex quantum circuits
that are easier to understand and implement.  A typical component
gate set contains a basis for Clifford operations together with a
single non-Clifford operation.  For brevity, we employ more Clifford
gates than are necessary to generate the Clifford group, augmenting
the gates of  Eqs.~(\ref{eq:HSdefinition}) and
(\ref{eq:CZdefinition}) by the Pauli matrices and $S^\dagger=S^3=ZS$.

Any state can be expressed in terms of a quantum circuit that
prepares it from some fiducial state, traditionally taken to be the
one in which each qubit is initially in the state $\ket{0}$.  To
prepare any $n$-qubit stabilizer state from $\ket{0}^{\otimes n}$, it
is sufficient to apply Clifford gates, since, by definition of the
Clifford group, there exists a Clifford operation that takes the
stabilizer of the fiducial state to the stabilizer of the desired
state under conjugation.  Applying this operation to the fiducial
state yields a $+1$ eigenstate of the stabilizers of the desired
state, i.e., the desired stabilizer state.

Graph states can be written in a particularly simple form using
quantum-circuit notation.  Any graph state can be prepared using two
layers of gates. In the first layer, $H$ is applied to each qubit.
In the second layer, $\CZ$ gates are applied between all pairs of
qubits corresponding to connected nodes on the graph.  To prepare an
arbitrary stabilizer state, it is sufficient to add a third layer
that contains only $H$ and $S$ gates.  This is because any stabilizer
state is equivalent to some graph state under local Clifford
operations~\cite{vandennest69}.

\section{Stabilizer-state graphs\label{sec:graphs}}

\subsection{Graphs from generator matrices\label{subsec:gengraphs}}

Any stabilizer state with a generator matrix of canonical form can be
converted by local Clifford operations to a state possessing a
generator matrix of strict graph form. Applying $H$ to the last $N-r$
qubits of the stabilizer state, where $r$ is the initial left rank of
the canonical-form generator matrix, exchanges columns $r+1$ through
$N$ in the generator matrix with columns $N+r+1$ through $2N$, so
that a generator matrix as in Eq.~(\ref{eq:canonicalstabilizer}) is
transformed to
\begin{equation}
\label{eq:intermediate}
\left(\begin{array}{cc|cc}I & 0 & B & A \\0 & I & A^T & 0\end{array}\right)\;.
\end{equation}
The diagonal of $B$ in this graph-form generator matrix can then be
stripped of $1$s without otherwise changing the generator matrix by
applying $S$ to offending qubits.  The resulting generator matrix has
the form of a graph state and corresponds to a stabilizer state that
differs from that represented by Eq.~(\ref{eq:canonicalstabilizer})
by at most a single $H$ or $S$ gate per qubit.

The close relationship between graph states and stabilizer states
suggests the possibility of a graph-like representation of stabilizer
states.  One approach to such a representation is simply to transform
the generator matrix of a stabilizer state into strict graph form,
draw the graph thereby obtained, and add decorations to each node
indicating whether an $H$ or $S$ was applied to the corresponding
qubit in the process of reaching strict graph form.  Thus are
stabilizer graphs constructed in this paper, where we choose to
signal the application of Hadamard gates by hollow (unfilled) nodes
and the application of phase gates by loops. The decision to
represent $S$ gates by loops is motivated by the standard graph
convention that a $1$ on the diagonal of an adjacency matrix denotes
a loop on that node.

\begin{figure}
\includegraphics[width=8.5cm]{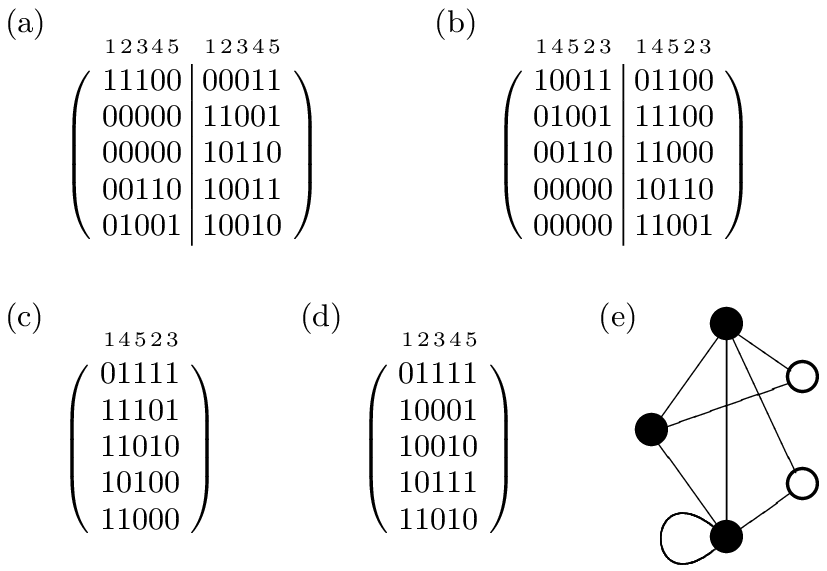}
\caption{(a)~A generator matrix for a stabilizer state. (b)~A
canonical-form generator matrix obtained from (a) by row and qubit
swapping. (c)~The adjacency matrix indicated by (b).  (d)~The
adjacency matrix of (c) with the qubit swaps undone.  (e)~The reduced
stabilizer graph associated with (a). In parts (a)-(d) the columns
have been labeled by the corresponding qubit.  In part (e) the nodes
are labeled sequentially, beginning with the top node and moving
clockwise. It is clear that the qubit swaps are not actually
necessary, since we reverse them in the end. The generator matrix in
(a) can be converted to graph form directly by exchanging columns $2$
and $3$ on the left with the matching columns on the right, an
operation that corresponds to applying a Hadamard to qubits $2$ and
$3$.  In graph form, the adjacency matrix is just the right half of
the generator matrix. Loops arise from $1$s on the diagonal of the
adjacency matrix, and hollow nodes are used to indicate which columns
were exchanged between the right and left halves of the generator
matrix to get the adjacency matrix. Notice that there are no edges
between hollow nodes, nor are there any loops on hollow
nodes.\label{fig:graph}}
\end{figure}

The following steps provide a recipe for translating an arbitrary
stabilizer generator matrix into a \textit{stabilizer graph}:

\begin{enumerate}
\item Through row reduction and qubit swapping, transform the
generator matrix into canonical form, as in
Eq.~(\ref{eq:canonicalstabilizer}), keeping track during the process
of how columns of the generator matrix map to qubits.

\item Draw the graph corresponding to the adjacency matrix
\begin{equation}
\label{eq:BAAT0}
\left(\begin{array}{cc}B&A\\A^T&0\end{array}\right)\;,
\end{equation}
including loops for $1$s on the diagonal.

\item Make solid the nodes corresponding to the rows and columns
of the submatrix $B$, and make hollow the nodes corresponding to the
rows and columns of the submatrix $0$.
\end{enumerate}

\noindent Notice that this procedure does not associate every
combination of edges, loops, hollow nodes, and solid nodes with a
stabilizer state. Because the submatrix $0$ in Eq.~(\ref{eq:BAAT0})
contains only $0$s, hollow nodes never have loops, and there are no
edges between hollow nodes.  We refer to stabilizer graphs having
this property as \textit{reduced}. An example of a generator matrix
and an associated reduced stabilizer graph is given in
Fig.~\ref{fig:graph}.

It is important to note that the graphs in this paper are
\textit{labeled graphs\/} in that each node is associated with a
particular qubit.  Swapping two qubits is a physical operation that
generally produces a different quantum state.  In our graphs a
\textit{SWAP\/} gate can be described either by relabeling the
corresponding nodes or by exchanging the nodes and all their
decorations and connections while leaving the labeling the same.
Since the process of bringing the generator matrix into canonical
form can involve swapping qubits, we must keep track during this
process of the correspondence between qubits and columns of the
generator matrix and thus between these columns and the nodes of our
graphs.

Just as a generator matrix contains no information about generator
signs, so also are stabilizer graphs derived from generator matrices
devoid of such information. It is for this reason that we can be
cavalier about whether $S$ or $S^\dagger=ZS$ is used to convert a
canonical generator matrix to strict graph form.  In the absence of
sign information, however, stabilizer graphs are best thought of as
specifying an orthonormal basis rather than a single stabilizer
state.  Luckily, it is not hard to include sign information in the
graph, as we show in the next subsection.

\subsection{Graphs from quantum circuits}
\label{subsec:circuitgraphs}

Having motivated a stabilizer-graph notation using generator
matrices, we now turn to the quantum-circuit formalism to expand it.
The binary representation of stabilizers lacks a convenient way to
keep track of generator signs, applied gates, and qubit swaps.
Since, for example, labeling is important, the column exchanges
required to bring a generator matrix to canonical form must be
tracked, perhaps by appending an extra row with qubit labels to the
generator matrix. Details such as this are automatically dealt with
when deriving stabilizer graphs from quantum circuits.

Consider a quantum circuit consisting of three layers of gates
applied to $N$ qubits, each initially in the state $\ket{0}$.  In the
first layer, the Hadamard gate, $H$, is applied to each qubit.  In
the second, controlled-sign gates, $\CZ$, are applied between various
pairs of qubits.  Finally, in the third layer, sign-flip gates, $Z$,
phase gates, $S$, and Hadamard gates are applied (in that order) to
various subsets of qubits.  We refer to such circuits as having
\textit{graph form}.

\begin{figure}
\center
\includegraphics[width=8.5cm]{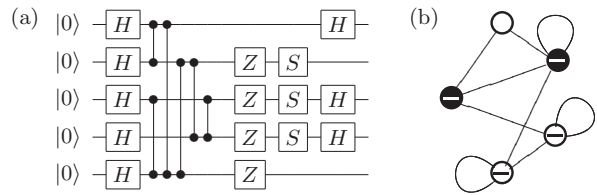}
\caption{(a)~A circuit in graph form. (b)~A stabilizer graph
corresponding to the circuit in (a). $\protect\CZ$ gates between
qubits are transformed into links between nodes, terminating $Z$
gates become negative signs, terminating $S$ gates result in loops,
and terminating $H$ gates are denoted by hollow nodes.  Nodes in~(b)
are labeled sequentially, beginning with the top node and moving
clockwise. \label{fig:stabgraph}}
\end{figure}

The descriptor arises because a quantum circuit in this form can be
depicted as a graph possessing three kinds of decoration. In such a
graph, two nodes are linked if a $\CZ$ gate is applied between the
qubits of the circuit corresponding to those nodes; the various kinds
of node decoration indicate the presence or absence of terminal $Z$,
$S$, and $H$ gates on the corresponding qubits.  It is the
restrictions of such a representation that impel us to specify an
order for the terminal gates, since $SH\neq HS$.  The decoration
corresponding to each gate is as follows: $Z$ gates are denoted by a
minus sign in the node, $S$ gates by a loop, and $H$ gates by a
hollow (as opposed to a solid) node.  We refer to an arbitrary
arrangement of solid and hollow nodes with loops, edges, and signs as
a \textit{stabilizer-state graph\/} or, more simply, as a
\textit{stabilizer graph}.    An example graph-form circuit and the
corresponding stabilizer graph are given in Fig.~\ref{fig:stabgraph}.

As might be guessed from our choice of decorations, the $Z$ gates in
a graph-form circuit specify the signs of the stabilizer generators.
The gates applied in the third layer of a graph-form circuit can be
written as
\begin{align}
\prod_{j=1}^N H_j^{c_j}S_j^{b_j}Z_j^{a_j}\;,
\label{eq:thirdLayer}
\end{align}
where $a_j$, $b_j$, and $c_j$ are binary variables taking on the
values $0$ or $1$. From Eq.~(\ref{eq:thirdLayer}) it follows that the
third layer of gates transforms a set of graph-state generators as in
Eq.~(\ref{eq:graphgenerators}) to the following stabilizer-state
generators:
\begin{align}
g_j=&(-1)^{a_j+b_jc_j}\nonumber\\
&\times X_j^{(b_j+1)(c_j+1)}Y_j^{b_j}Z_j^{(b_j+1)c_j}
\prod_{k\in\mathcal{N}(j)}Z_k^{c_k+1}X_k^{c_k}\;.
\label{eq:stabilizerStateGenerators}
\end{align}
Equation~(\ref{eq:stabilizerStateGenerators}) shows that the
exclusive effect of each terminal $Z$ operator is to flip the sign of
a single stabilizer generator.  This can also be seen through circuit
identities, since pushing a $Z$ gate from the third layer of a
graph-form quantum circuit to the beginning of the circuit merely
transforms it to an $X$ gate; flipping an input bit is equivalent to
flipping the sign of the associated stabilizer since the stabilizer
of $\ket{0}$ is $Z$, which acquires a negative sign under conjugation
by $X$.  Using either of these methods, it is clear that terminal $Z$
gates, and hence the signs in stabilizer graphs, only impact the
signs of the stabilizer generators, and can thus be omitted when
these signs are thought to be unimportant.

Modulo generator signs, the definition of stabilizer graphs given in
the previous subsection and the definition given in this one are
compatible.  The graph-form circuit and generator matrix associated
with a stabilizer graph each specify the same stabilizer up to
possible signs.  The method of deriving stabilizer graphs from
generator matrices described in Sec.~\ref{subsec:gengraphs}, however,
produces exclusively reduced stabilizer graphs, i.e., stabilizer
graphs satisfying the restriction that hollow nodes never have loops
and never be connected to other hollow nodes.  In terms of graph-form
quantum circuits, this is the restriction that lines with terminating
Hadamard gates have no terminating $S$ gates and not be connected by
$\CZ$ gates.  The relative merits of stabilizer graphs and reduced
stabilizer graphs are clarified in the following sections,
particularly Sec.~\ref{sec:equiv}.  Important roles are found for
both.

\begin{figure*}
\center
\includegraphics[width=18cm]{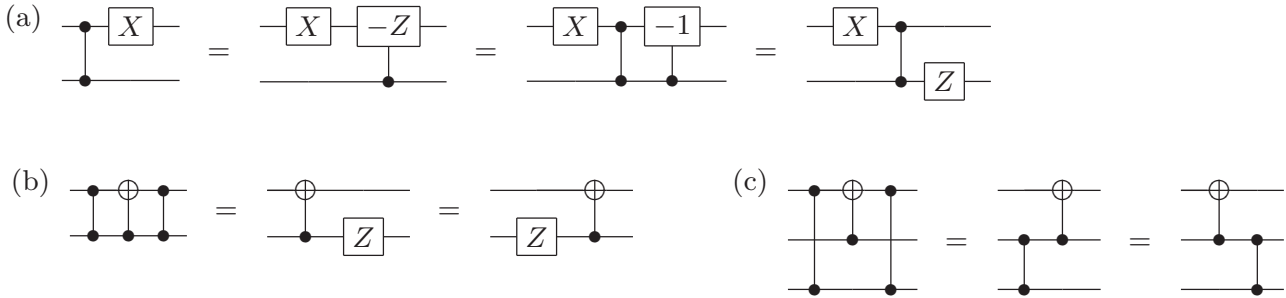}
\caption{Simple circuit identities, which we use to derive more
complex identities.  Identity~(a) follows trivially from the steps
shown. Identities~(b) and (c) are easily verified in the standard
basis. \label{fig:basics}}
\end{figure*}

\section{Graph transformations}
\label{sec:transformations}

It is frequently useful to consider the way in which stabilizer
states transform under the application of Clifford gates.  Primarily,
this is because Clifford gates take stabilizer states to stabilizer
states, a property that follows from their preservation of the Pauli
group.  This same property implies that the action of a Clifford gate
can be thought of as a transformation between the graphs representing
the initial and final stabilizer states.  In this section we consider
the transformations induced by local Clifford gates.  The
transformations induced by $\CZ$ gates are discussed separately in the
Appendix.

\subsection{Terminology}
\label{subsec:terminology}

We begin by introducing terminology for describing transformations on
stabilizer graphs.   We use a number of terms, some adopted from
graph theory and others invented for the task at hand.

Among those terms common to graph theory are \textit{neighbors},
\textit{complement}, and \textit{local complement}.  The neighbors of
a node~$j$ are those nodes connected to $j$ by edges; we denote the
set of neighbors of node~$j$ by $\mathcal{N}(j)$.  In the
transformation rules that follow, a loop does not count as an edge,
so a node is never its own neighbor. Complementing the edge between
two nodes removes the edge if one is present and adds one otherwise.
The local complement is performed by complementing a selection of
edges, with the pattern of edges depending on whether local
complementation is applied to a node or along an edge.

Performing local complementation on a node complements the edges
between all of the node's neighbors.  Thus, local complementation at
node~$j$ transforms the adjacency matrix~$\Gamma$ of a graph
according to
\begin{equation}
\label{eq:lc}
\Gamma_{lm}\rightarrow\Gamma'_{lm}=
\Gamma_{lm}+(1+\delta_{lm})\Gamma_{jl}\Gamma_{jm}\;;
\end{equation}
i.e., it complements an edge if both nodes of the edge
are neighbors of $j$.

Local complementation along an edge is equivalent to a sequence of
local complementations on the \textit{decision nodes\/}, i.e., the
nodes defining the edge.  The sequence is as follows: first perform
local complementation on one of the decision nodes, then local
complement on the other decision node, and finally local complement
on the first decision node again.  This sequence transforms the
adjacency matrix of a simple graph in the following way:
\begin{eqnarray}
\Gamma_{lm}\rightarrow\Gamma'_{lm}&=&\Gamma_{lm}
+(\Gamma_{jl}+\delta_{jl})(\Gamma_{km}+\delta_{km})\nonumber\\
&&\quad{}+(\Gamma_{jm}+\delta_{jm})(\Gamma_{kl}+\delta_{kl})\;,
\label{eq:threelc}
\end{eqnarray}
where $j$ and $k$ are the decision nodes.
Equation~(\ref{eq:threelc}) is symmetric in the two decision nodes,
so it does not matter at which decision node local complementation is
first performed. Additionally, since we do not consider self-loops to
be edges, Eq.~(\ref{eq:threelc}) can be applied to adjacency matrices
with nonzero diagonal entries simply by ignoring those entries.
Notice that an edge one of whose nodes is a decision node transforms
according to $\Gamma'_{lj}=\Gamma_{lk}$.

To describe the net effect of local complementation along an edge, it
is helpful to define the \textit{decision neighborhood\/} of a node
as the intersection of its neighborhood with the decision nodes.
Then local complementation along an edge can be summarized by three
steps:

\begin{enumerate}

\item The edge between the decision nodes is left unchanged.

\item An edge one of whose nodes is a decision node is transferred
from this decision node to the other.

\item An edge neither of whose nodes is a decision node is
complemented if its end nodes have decision neighborhoods that
are not empty and not identical.

\end{enumerate}

To these terms we add \textit{flip\/} and \textit{advance}. Flip is
used to describe the simple reversal of some binary property, such as
the sign or the fill state (color) of a node.  Advance refers
specifically to an action on loops; advancing generates a loop on
nodes where there was not previously one, and it removes the loop and
flips the sign on nodes where there was a loop.  Its action mirrors
the application of the phase gate, since $S^2=Z$.

\begin{figure*}
\center
\includegraphics[width=18cm]{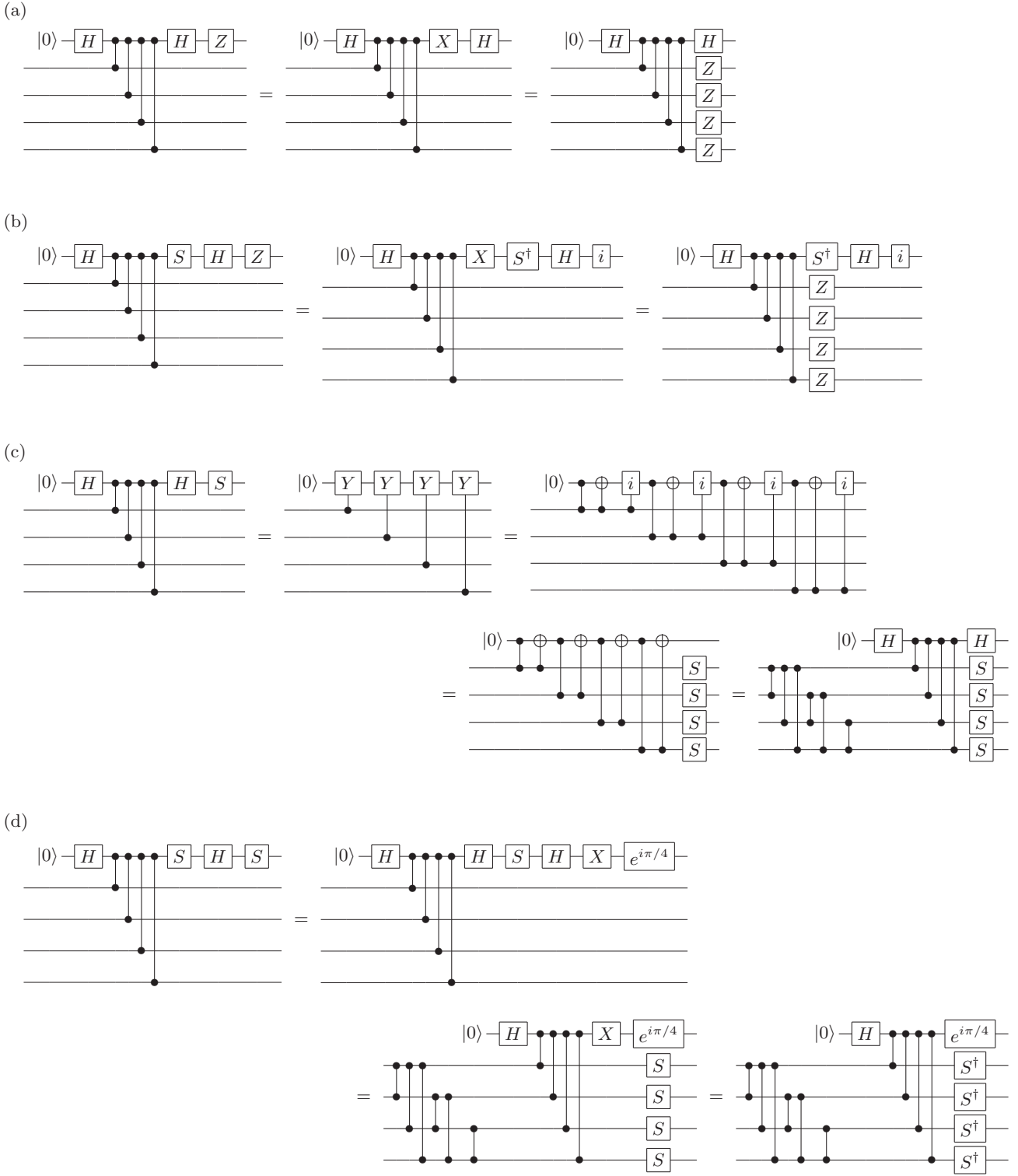}
\caption{Circuit identities relevant to transforming a stabilizer
graph under local Clifford operations.  Each identity is illustrated
for the case of four neighbors connected to a qubit of interest;
equivalent expressions hold for other numbers of neighbors.
Identities~(a) and~(b) follow immediately from the identity in
Fig.~\ref{fig:basics}(a).  The proof of~(c) relies on decomposing
$\protect\C{Y}$ using $Y=iXZ$, the equivalence of $\protect\C{i}$ to
$S$ on the control qubit, and the identity in
Fig.~\ref{fig:basics}(c).  The first step in identity~(d) uses the
fact that $I=e^{-i\pi/4}(HS)^3$, which gives $SHS=e^{i\pi/4}XHSH$;
the second and third steps follow from identities~(c) and~(a). The
identity in Fig.~\ref{fig:basics}(c) generates the thicket of
controlled-sign gates in the lower left of the final circuits in (c)
and (d), thereby giving rise to local complementation in the
corresponding stabilizer graphs.  In~(b) and~(d), the phase shifts on
the top qubits, $i$ and $e^{i\pi/4}$, are global phase shifts and
thus can be omitted.  To complete the transformation rules, it is
necessary to know what happens in each of these circuit identities
when there is a $Z$ gate on the top qubit immediately after the
$\protect\C{Z}$ gates.  Ignoring overall phases, the effect on the
third layer of the final circuit is, for~(a) and~(b), to include an
additional $Z$ on the top qubit and, for~(c) and~(d), to include an
additional $Z$ on all qubits. \label{fig:identproof}}
\end{figure*}

\subsection{Stabilizer-graph transformations\label{subsec:stabgraphtrans}}

An arbitrary local Clifford operation can be decomposed into a
product of $H$, $S$, and $Z$ gates (the $Z$ is unnecessary,
but convenient). The effect of a local Clifford operation on a
stabilizer graph can thus be obtained by repeated application of the
following six transformation rules.

\begin{enumerate}
\item[T1.] Applying $H$ to a node flips its fill.
\item[T2.] Applying $S$ to a solid node advances its loop.
\item[T3.] Applying $S$ to a hollow node without a loop performs
local complementation on the node and advances the loops of its
neighbors.

If the node has a negative sign, flip the signs of its neighbors as
well.
\item[T4.] Applying $S$ to a hollow node with a loop flips its
fill, removes its loop, performs local complementation on it, and
advances the loops of its neighbors.

If the node does not have a negative sign, flip the signs of its
neighbors as well.
\item[T5.] Applying $Z$ to a solid node flips its sign.
\item[T6.] Applying $Z$ to a hollow node flips the signs of all of
its neighbors.  If the node has a loop, its own sign is flipped as
well.
\end{enumerate}

These transformation rules can be derived from the circuit identities
in Fig.~\ref{fig:identproof}, which rely on the basic circuit
identities given in Fig.~\ref{fig:basics}.   Given an understanding
of the relationship between circuits and graphs, transformation rules
T1, T2, and T5 are trivial. Transformation rule T6 follows simply
from Fig.~\ref{fig:identproof}(a,b). Transformation rules T3 and T4
derive from Figs.~\ref{fig:identproof}(c) and (d), respectively.

\subsection{Reduced-stabilizer-graph transformations\label{subsec:redstabgraphtrans}}

The transformation rules T1--T6 do not generally take reduced
stabilizer graphs to reduced stabilizer graphs.  From
Sec.~\ref{subsec:gengraphs}, however, we know that there exists a
reduced stabilizer graph corresponding to each stabilizer state, so
it is always possible to represent the effect of a local Clifford
operation as a mapping between reduced stabilizer graphs.  The
appropriate transformation rules for reduced stabilizer graphs are
listed below, excepting those for $Z$ operations, which are identical
to T5 and T6.

\begin{list}{}{\itemindent=0pt\leftmargin=34pt\labelsep=5pt\labelwidth=34pt}
\item[T(i).] Applying $H$ to a solid node without a loop, which is
only connected to other solid nodes, flips the fill of that node.

\item[T(ii).] Applying $H$ to a solid node with a loop, which is only
connected to other solid nodes, performs local complementation on the
node and advances the loops of its neighbors.

Flip the node's sign, and if it now has a negative sign, flip the
signs of its neighbors as well.
\item[T(iii).] Applying $H$ to a solid node without a loop, which is
connected to a hollow node, flips the fill of the hollow node and
performs local complementation along the edge connecting the nodes.

Flip the sign of nodes connected to both the solid and hollow nodes.
If either of these two nodes has a negative sign, flip it and
the signs of its current neighbors.

\item[T(iv).] Applying $H$ to a solid node with a loop, which
is connected to a hollow node, performs local complementation on the
solid node and then on the hollow node.  Then it removes the loop from
the solid node, advances the loops of the solid node's current
neighbors, and flips the fill of the hollow node.

Flip the signs of nodes that were originally connected to both the
solid and hollow nodes.  If the originally solid node initially had a
negative sign, flip it and the signs of its current neighbors, and if
the originally hollow node initially had a negative sign, flip the
signs of its current neighbors.

\item[T(v).] Applying $H$ to a hollow node flips its fill.
\item[T(vi).] Applying $S$ to a solid node advances its loop.
\item[T(vii).] Applying $S$ to a hollow node performs local
complementation on that node and advances the loops of its neighbors.

If the node has a negative sign, flip the signs of its neighbors as well.
\end{list}

Of these transformation rules, T(i), T(v), and T(vi) are trivial, and
T(vii) is a rewrite of T3.   To prove the others requires results
from Sec.~\ref{sec:equiv}, in particular, the equivalence rules in
Sec.~\ref{subsec:circuitequiv}.  Specifically, T(ii) is obtained by
applying equivalence rule~E1, which gives an equivalent, but
unreduced graph and then applying the Hadamard, via rule~T1, which
leaves a reduced graph.  For T(iii), one first applies the Hadamard,
via rule~T1, and then uses equivalence rule~E2 to convert to a
reduced graph.  In the case of T(iv), one applies the Hadamard, using
rule~T1, and then applies equivalence rule E1, first to the
originally solid node and then to the hollow node.  A key part of
these transformations is the conversion of stabilizer graphs to
reduced form, a process discussed in more detail in
Sec.~\ref{subsubsec:convert}.

It is not hard to check that, in using rules T(iii) and T(iv), any
other hollow nodes that are connected to the originally solid node do
not become connected and do not acquire loops, in accordance with the
need to end up with a reduced graph.

\section{Equivalent stabilizer graphs\label{sec:equiv}}

As we have defined it, the mapping between a graph-form circuit and
its corresponding stabilizer graph is one-to-one.  This does not
imply, however, that each stabilizer state corresponds to a unique
graph.  On the contrary, an example of different graph-form circuits
corresponding to the same stabilizer state can be found in
Fig.~\ref{fig:identproof}.  Applying an additional $S$ gate to the
top qubit in Fig.~\ref{fig:identproof}(d) makes the initial circuit
identical to that in~(b), but the final circuits are quite different.
The two circuit identities thus define distinct transformation rules
for applying a $Z$ gate to a hollow node with a loop, thereby
demonstrating the existence of multiple, equivalent graph-form
circuits. For every way of obtaining a particular stabilizer state
from a circuit in graph form, there is an associated stabilizer
graph. In this section we examine the resulting equivalence classes
of stabilizer graphs, first by presenting equivalence rules for full
and reduced stabilizer graphs and then by introducing simplified
graph-form-circuit equalities which we use to show that the
equivalence rules given here are complete.

\begin{figure*}
\center
\includegraphics[width=18cm]{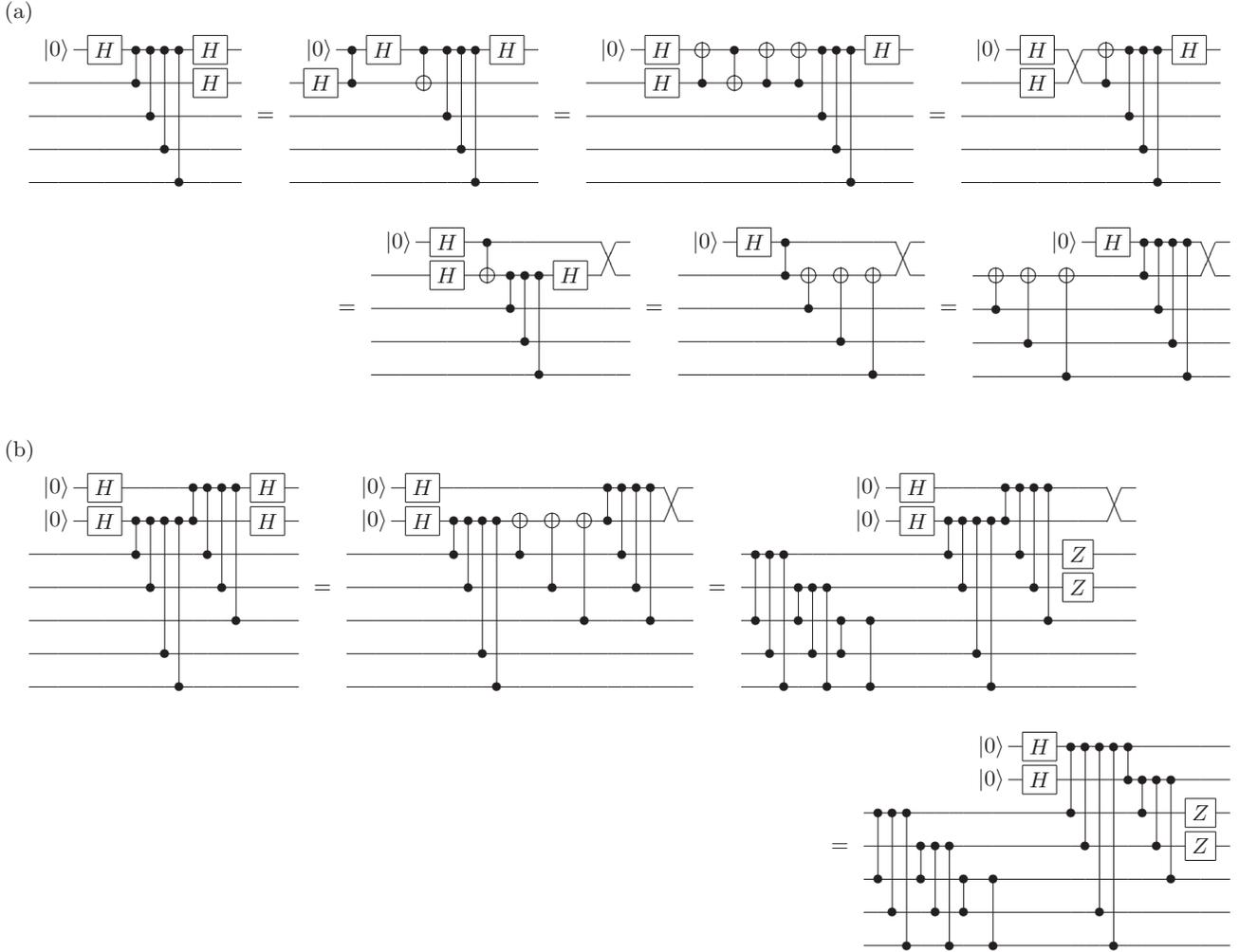}
\caption{(a) An illustration of a circuit identity used in the
removal of pairs of Hadamards.  The identity is shown for the case of
four neighbors connected to a qubit of interest; equivalent
expressions hold for other numbers of neighbors.  The crucial trick
here is the use of the decomposition of \textit{SWAP} in terms of
three $\protect\CX$ gates. (b) A circuit identity demonstrating
transformation rule~E2.  The selection of neighbors of the top two
qubits, the decision qubits, is chosen to display the behavior of all
kinds of edges in which both end nodes are connected to at least one
of the decision qubits.  The first equality in~(b) employs the
identity from part~(a), while the subsequent ones follow from basic
circuit identities.  Concerning edges, the net effect of E2 on a
stabilizer graph is to perform local complementation along the edge
joining the decision nodes, i.e., to swap the connections of the
decision nodes and to complement any edges whose end nodes connect to
the decision nodes in different ways.  The latter of these effects is
contained in the controlled-sign gates in the lower left of the final
circuit.  Circuits illustrating E2 for other initial fill states can
be obtained simply by applying additional terminal Hadamards to the
circuits in~(b).  The sign rule for E2 expresses the effect of a $Z$
gate on a decision qubit before the Hadamard gate in the third
layer of the initial circuit.  Such a $Z$ gate can be pushed through
the succeeding Hadamard, becoming an $X$ gate; in the final circuit,
this $X$ gate, when processed using the identity in
Fig.~\ref{fig:basics}(a), deposits an additional $Z$ gate on the
qubits that were originally connected to the other decision qubit,
excluding the decision qubit that originally possessed the $Z$.
\label{fig:transIdent}}
\end{figure*}

\subsection{Stabilizer-graph equivalences}
\label{subsec:circuitequiv}

Applying either of the following two rules to a stabilizer graph
yields an equivalent graph, i.e., one which represents the same
stabilizer state.

\begin{enumerate}
\item[E1.] Flip the fill of a node with a loop.  Perform local
complementation on the node, and advance the loops of its neighbors.

Flip the node's sign, and if the node now has a negative sign, flip
the signs of its neighbors as well.

\item[E2.] Flip the fills of two connected nodes without loops, and
local complement along the edge between them.

Flip the signs of nodes connected to both of the two original nodes.
If either of the two original nodes has a negative sign, flip it and
the signs of its current neighbors.
\end{enumerate}

The first of these equivalence rules can be obtained by applying an
additional $S$ gate to the top qubit in the identity of
Fig.~\ref{fig:identproof}(d) and equating the final circuit to the
final circuit in Fig.~\ref{fig:identproof}(b).  For the second rule,
we need yet another circuit identity.  Figure~\ref{fig:transIdent}(a)
shows how Hadamards can be removed from a pair of connected qubits
without $S$ gates.  Figure~\ref{fig:transIdent}(b) extends this
identity to a demonstration of rule~E2.

\subsection{Reduced-stabilizer-graph equivalences}
\label{subsec:redequiv}

The equivalence rules of the previous section can be reworked to
yield equivalence rules for reduced stabilizer graphs.  The resulting
equivalence rules are

\begin{list}{}{\itemindent=0pt\leftmargin=30pt\labelsep=5pt\labelwidth=30pt}

\item[E(i).]For a hollow node connected to a solid node with a
loop, local complement on the solid node and then on the hollow node.
Then remove the loop from the solid node, advance the loops of its
current neighbors, and flip the fills of both nodes.

As for signs, follow the sequence in transformation rule T(iv).
\item[E(ii).]For a hollow node connected to a solid node without
a loop, local complement along the edge between them.  Then flip the
fills of both nodes.

As for signs, follow the sequence in equivalence rule~E2.
\end{list}

There is a simple relationship between the two sets of equivalence
rules.  Equivalence rule E(ii) is identical to E2 for the case that
the two connected nodes have opposite fill.  Equivalence rule E(i) is
simply rule E1 applied twice: first to the solid node with the loop
and then once to the hollow node that has acquired a loop from the
first application of E1.  The second application of E1 is needed
because the resulting graph is not reduced after one employment of
the rule.

Both of these equivalence rules can also be derived by applying two
Hadamards to a solid node, E(i) handling the case in which the solid
node has a loop and E(ii) the case in which it does not. Thus
equivalence rule~E(i) is simply transformation rule~T(iv) followed by
use of rule~T(i) to apply a second Hadamard to the originally solid
node. Likewise, E(ii) is rule~T(iii) followed by T(i) to apply a
second Hadamard to the originally solid node.  Notice that both rules
preserve the number of hollow nodes.

\begin{figure*}
\center
\includegraphics[width=18cm]{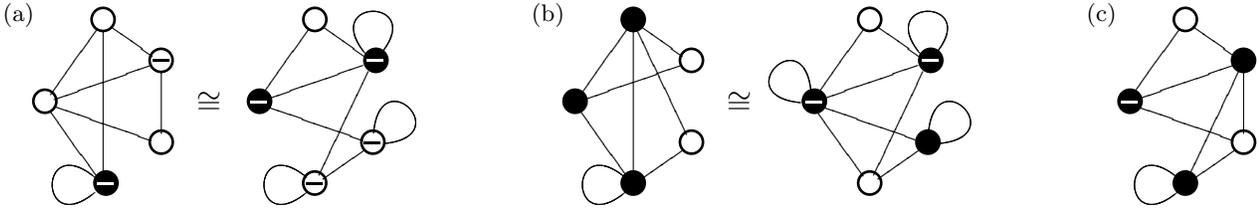}
\caption{An example sequence of graphs representing parts~$1$ through
$3$ of the equivalence-checking procedure in
Sec.~\ref{subsec:equivalenceproof}.  Nodes are labeled sequentially,
beginning at the top node and moving clockwise.  The graph on the
left-hand side of~(a) is equivalent to the reduced graph on the
left-hand side of~(b) by the application of~E2 to the pair of
nodes~$\{1,5\}$.  Likewise, an application of~E1 to node~$3$ in the
graph on the right-hand side of~(a) yields the reduced graph on the
right-hand side of~(b).  The graph of~(c) results from the
application of~E(ii) to the pair of nodes~$\{1,2\}$ in the graph on
the left of~(b).  An application of~E(i) to the pair of
nodes~$\{3,4\}$ in the graph on the right-hand side of~(b) also
yields the graph in~(c), verifying, as per
Sec.~\ref{subsubsec:evaluate}, that the graphs in~(a) represent the
same state.\label{fig:equivalenceexample}}
\end{figure*}

\subsection{Constructive procedure for showing sufficiency of equivalence rules}
\label{subsec:equivalenceproof}

Having described a set of rules for converting between equivalent
stabilizer graphs, we show in this section that, in each case, the
aforementioned rules generate the entire equivalence class of
stabilizer graphs.  The proof is divided into three parts.  The first
part shows how to use rules~E1 and E2 to convert an arbitrary
stabilizer graph to an equivalent graph in reduced form.  The second
part explains how an equivalence test for a pair of reduced
stabilizer graphs can be simplified, using rules~E(i) and~E(ii), to a
special form.  Finally, the third part proves that the graphs on the
two sides of such a simplified equivalence test are equivalent only
if they are trivially identical.  Taken as a whole, this proves that
the set of equivalence rules given in Secs.~\ref{subsec:circuitequiv}
and \ref{subsec:redequiv} is sufficient to convert (reversibly)
between any two equivalent stabilizer graphs and thus that they
generate all stabilizer graphs that are equivalent to the same
stabilizer state. Similarly, considering only the final two parts of
the proof shows that the rules given in Sec.~\ref{subsec:redequiv}
are sufficient to generate all reduced stabilizer graphs.

\subsubsection{Converting stabilizer graphs to reduced form}
\label{subsubsec:convert}

Two features identify a stabilizer graph as reduced.  In a reduced
graph, hollow nodes never have loops, and hollow nodes are never
connected to each other.  Equivalence rule~E1 can be used to
convert looped nodes from hollow to solid.  Applying rule E1 in
this sort of situation can cause hollow nodes to acquire loops, but
each application fills one hollow node of the graph, so the procedure
will terminate in at most a number of repetitions equal to the number
of hollow nodes in the graph.  Similarly, connected hollow nodes in
the resulting graph can be converted to solid nodes using the
appropriate case of rule E2.  Once again, the process is guaranteed
to terminate because the number of hollow nodes to which the rule
might be applied decreases by two with each application of the rule.
Concomitantly, the conversion of a stabilizer graph to an equivalent
reduced graph never increases the number of hollow nodes.

\subsubsection{Simplifying reduced-graph equivalence testing}
\label{subsubsec:simplify}

Equivalence testing for pairs of reduced graphs is facilitated by
simplifying the equivalence such that nodes that are hollow only in
the first graph never connect to nodes that are hollow only in the
second.  This simplification can be accomplished by iterating the
following process.  Choose a pair of connected (in either graph)
nodes $a$ and $b$ such that $a$ is hollow in one graph and $b$ is
hollow in the other, and to the graph in which they are connected,
apply the relevant reduced equivalence rule to the selected nodes.
Among other things, the equivalence operation reverses the fill of
the two nodes it is applied to.  Since one node is hollow and the
other solid, this preserves the total number of hollow nodes while
yielding a node that is hollow in both graphs.
Because it is applied only to unpaired hollow nodes, subsequent uses of this equivalence rule do not disturb the newly paired hollow node.
Consequently, this process also terminates in at most a number of
repetitions equal to the number of hollow nodes in each of the
graphs.

\subsubsection{Trivial evaluation of simplified reduced-graph\\equivalence tests}
\label{subsubsec:evaluate}

The two reduced graphs composing a simplified reduced-graph
equivalence test are equivalent, i.e., correspond to the same state,
if and only if the graphs are identical.  To see why this is so, we
return to the graph-form quantum circuits discussed earlier.  In
addition to the standard restrictions for reduced graphs, the
circuits corresponding to the graphs in a simplified equivalence test
have the following property: if in one of the circuits, a qubit with
a terminal $H$ participates in a $\CZ$ gate with a second qubit
(which cannot have a terminal $H$), then in the other circuit, it
cannot be true that the second qubit has a terminal $H$ and the first
does not.  We prove the triviality of simplified reduced-graph
equivalence testing by considering an arbitrary simplified
reduced-graph equivalence test and showing that the two graphs must
be identical if they are to correspond to the same state.

In terms of unitaries, an arbitrary graph-form circuit equality can
be written as
\begin{align}
  \begin{split}
\prod_{g\in\set{H}_l} &H_g
\prod_{j\in\set{S}_l} S_j
\prod_{h\in\set{Z}_l} Z_h
\prod_{\gamma\in\set{C}_l} \CZ_\gamma
\prod_{k} H_k \ket{0}^{\otimes n} \\
=&
\prod_{g\in\set{H}_r} H_g
\prod_{j\in\set{S}_r} S_j
\prod_{h\in\set{Z}_r} Z_h
\prod_{\gamma\in\set{C}_r} \CZ_\gamma
\prod_{k} H_k \ket{0}^{\otimes n}\;,
  \end{split}
  \label{eq:arbSimpGraphFormCircEqual}
\end{align}
where $\set{C}$ lists the pairs of qubits participating in $\CZ$
gates and $\set{H}$, $\set{Z}$, and $\set{S}$ are sets enumerating
the qubits to which $H$, $Z$, and $S$ gates are applied respectively.
The total number of qubits is denoted by $n$ and the subscripts $l$
and $r$ discriminate between the circuits on the left- and right-hand
sides of the equation. In terms of these sets, a reduced-graph-form
circuit satisfies the constraints $\set{H}\cap\set{S}=\varnothing$
and $\{a,b\}\not\in\set{C}$ for all $a,b\in\set{H}$.  The circuits in
simplified tests also satisfy $\{a,b\}\not\in\set{C}_l,\set{C}_r$ for
all $a\in\hhbar$ and $b\in\hbarh$, where $\bar{\set{H}}$ denotes the
complement of $\set{H}$, i.e., the set of qubits to which $H$ is
not applied.

Suppose now that the two graphs have hollow nodes at different
locations; i.e., at least one of the sets, $\hhbar$ and $\hbarh$, is
not empty.  For specificity, let's say that $\hhbar$ is not empty. In
the language of circuits, this means that there exists a qubit $a$
that has a terminal $H$ on the left side of
Eq.~(\ref{eq:arbSimpGraphFormCircEqual}), but not on the right side.
Since qubit~$a$ is part of a circuit for a reduced graph, it does not
participate in $\CZ$ gates with other qubits that possess terminal
$H$ gates.  Consequently, on the left side of
Eq.~(\ref{eq:arbSimpGraphFormCircEqual}), the $\CZ$ gates involving
qubit $a$ can all be moved to the end of the circuit where they
become $\CX$ gates with $a$ as the target.  Doing this and
transferring the $\CX$ gates to the other side yields,
\begin{widetext}
\begin{align}
  \begin{split}
\prod_{g\in\set{H}_l} H_g
\prod_{j\in\set{S}_l} S_j
\prod_{h\in\set{Z}_l} Z_h
\prod_{\gamma\in\set{C}_l\mbox{\scriptsize{\ s.t.\ }}a\not\in\gamma}\!\!\!\!\!\CZ_\gamma\;\;\prod_{k} H_k
\ket{0}^{\otimes n} =\prod_{b\in\set{N}_l(a)} \C{X}_{ba}
\prod_{g\in\set{H}_r} H_g
\prod_{j\in\set{S}_r} S_j
\prod_{h\in\set{Z}_r} Z_h
\prod_{\gamma\in\set{C}_r} \CZ_\gamma
\prod_{k} H_k \ket{0}^{\otimes n}\;,
  \end{split}\label{eq:modCircEqual}
\end{align}
where $\set{N}_l(a)$ denotes the set of qubits that participate in
$\CZ$ gates with qubit $a$ on the left-hand side of
Eq.~(\ref{eq:arbSimpGraphFormCircEqual}).

Because the original graph
equality~(\ref{eq:arbSimpGraphFormCircEqual}) was simplified,
$\set{N}_l(a)\cap\set{H}_r=\varnothing$ and, by assumption,
$a\not\in\set{H}_r$, so the $\CX$ gates and the terminal Hadamards on
the right side of Eq.~(\ref{eq:modCircEqual}) do not act on the same
qubits.  Thus we can commute the $\CX$ gates past the terminal
Hadamards.  Moreover, we can then move the $\CX$ gates to the
beginning of the circuit where they have no effect and can therefore
be dropped.  During this migration, however, they generate a
complicated menagerie of phases.  The resulting expression for the
right side of Eq.~(\ref{eq:modCircEqual}) is
\begin{align}
  \begin{split}
\prod_{g\in\set{H}_r} H_g
\prod_{j\in\set{S}_r} S_j
\raisebox{-.35em}{\Bigg(\Bigg.}
\prod_{d\in\set{N}_{l}(a)}S_d \CZ_{da}
\raisebox{-.35em}{\Bigg.\Bigg)}^{\!\!\mathbf{1}_{\set{S}_r}(a)}
\prod_{h\in\set{Z}_r} Z_h
\raisebox{-.35em}{\Bigg(\Bigg.}
\prod_{f\in\set{N}_{l}(a)} Z_f
\raisebox{-.35em}{\Bigg.\Bigg)}^{\!\!\mathbf{1}_{\set{Z}_r}(a)}
\prod_{\gamma\in\set{C}_r}\CZ_\gamma
\prod_{\delta\in\set{N}(a)}\CZ_\delta
\prod_{c\in\set{N}_l(a)\cap\set{N}_r(a)}\!\!\!Z_c\;\;\prod_{k} H_k\ket{0}^{\otimes n}\;,
  \end{split}\label{eq:modCircEqualRight}
\end{align}
\end{widetext}
where $\set{N}_r(a)$ is defined similarly to $\set{N}_l(a)$,
\begin{equation}
\set{N}(a) = \{ \{ p, q \} | p\in\set{N}_l(a), q\in\set{N}_r(a),
p\neq q \}\;,
\end{equation}
and $\mathbf{1}$ represents an indicator function, e.g.,
$\mathbf{1}_{\set{Z}_r}(a)$ equals $1$ if $a\in\set{Z}_r$ and $0$
otherwise.

\begin{figure*}
\center
\includegraphics[width=18cm]{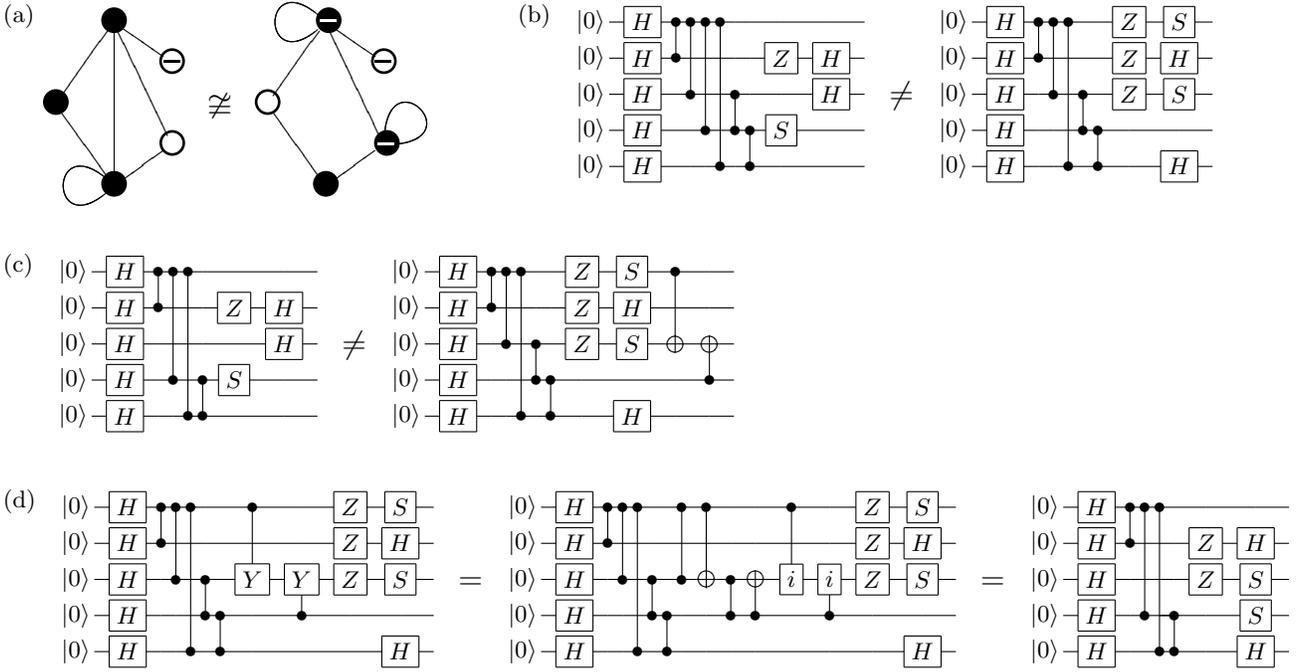}
\caption{ An example of the process described in
Sec.~\ref{subsubsec:evaluate}.  A simplified reduced-graph
equivalence test is shown in (a). The test is simplified because a
node that is hollow in only one graph is never connected to a node
that is hollow only in the other. The test is also false, since such
an equivalence is satisfied if and only if the two graphs are
identical.  By switching to circuit notation, it is possible to see
why these graphs correspond to different states. Translating the
graph equality in~(a) into circuit notation yields the circuit
equality in~(b), where the nodes of~(a) are taken to be numbered
sequentially, beginning at the top node and moving clockwise.  The
fact that the graphs are reduced implies that qubits with terminal
$H$ gates, such as qubit $3$ in the left-hand circuit, are not
connected to other qubits with terminal $H$ gates.  This allows us to
pull out the $\protect\CZ$ gates acting on qubit $3$ on the left side
and transfer the resulting $\protect\CX$ gates to the other side of
the equation, yielding the equality in~(c).  That the graph equality
was simplified guarantees that $H$ gates do not prevent us from then
pushing the $\protect\CX$ gates to the beginning of the right-hand
circuit in~(c), as shown in~(d).  Qubit 3 winds up separable on both
sides of the equation (though this is not generally the case),
allowing us to verify the inequivalence of the prepared states since
$\ket{0}\ne(\ket{0}-i\ket{1})/\sqrt2$. \label{fig:partthreeexample}}
\end{figure*}

It might appear that we have made things substantially worse by this
rearrangement, but in an important sense,
Eq.~(\ref{eq:modCircEqualRight}) is now very simple with regard to
qubit~$a$: $H$ is applied to qubit~$a$ followed by a sequence of
unitary gates all of which are diagonal in the standard basis.  This
implies that there are an equal number of terms in the resultant
state where qubit~$a$ is in the state $\ket{0}$ and the state
$\ket{1}$.  On the left side of Eq.~(\ref{eq:modCircEqual}), however,
the only gate remaining on qubit~$a$ is the identity or an $X$,
depending on whether $a\in\set{Z}_l$; thus qubit~$a$ is separable and
is either in state $\ket{0}$ or $\ket{1}$ depending on whether
$a\in\set{Z}_l$.  Consequently, our initial assumption that
$\set{H}_l \neq \set{H}_r$ is incompatible with satisfying the
equality.

The preceding discussion shows that two graphs composing a simplified
equivalence test are equivalent only if they have hollow nodes in
exactly the same locations.  Given this constraint, the terminal
Hadamards can be canceled from both sides of
Eq.~(\ref{eq:arbSimpGraphFormCircEqual}), giving
\begin{align}
\begin{split}
\prod_{h\in\set{Z}_l}& Z_h \prod_{j\in\set{S}_l} S_j \prod_{\gamma\in\set{C}_l} \CZ_\gamma
\prod_{k} H_k \ket{0}^{\otimes n} \\
=&\prod_{h\in\set{Z}_r} Z_h \prod_{j\in\set{S}_r} S_j \prod_{\gamma\in\set{C}_r} \CZ_\gamma
\prod_{k} H_k \ket{0}^{\otimes n}\;.\
\end{split}\label{eq:paredSimpGraphFormCircEqual}
\end{align}
The state after the initial Hadamards is an equally weighted
superposition of all the states in the standard basis.  The
subsequent unitaries are diagonal in the standard basis, so they put
various phases in front of the terms in the equal superposition.
Since a unitary is fully described by its action on a complete set of
basis states, demanding equality term-by-term in
Eq.~(\ref{eq:paredSimpGraphFormCircEqual}) amounts to requiring that
\begin{align}
\prod_{h\in\set{Z}_l}& Z_h
\prod_{j\in\set{S}_l} S_j
\prod_{\gamma\in\set{C}_l} \CZ_\gamma=
\prod_{h\in\set{Z}_r} Z_h
\prod_{j\in\set{S}_r} S_j
\prod_{\gamma\in\set{C}_r} \CZ_\gamma\;,
\end{align}
which is only satisfied when $\set{Z}_l=\set{Z}_r$,
$\set{S}_l=\set{S}_r$, and $\set{C}_l=\set{C}_r$.  Thus, after
simplification, the equivalence of pairs of reduced graphs is trivial
to evaluate, since equivalence requires that the two graphs be
identical.

An example of the entire process of testing graph equivalence is
given in Fig.~\ref{fig:equivalenceexample}. An example which
illustrates the circuit manipulations described algebraically in the
text of this section is given in Fig.~\ref{fig:partthreeexample}.

As mentioned above, our proof provides a constructive procedure for
testing the equivalence of stabilizer graphs.  Moreover, it shows
that the set of equivalence rules given in
Sec.~\ref{subsec:circuitequiv} is sufficient to convert between any
equivalent stabilizer graphs and that the rules given in
Sec.~\ref{subsec:redequiv} are sufficient to convert between any
equivalent reduced stabilizer graphs.  Since the conversion of an
arbitrary stabilizer graph to reduced form never increases the number
of hollow nodes and the rules~E(i) and E(ii) that convert among
reduced graphs preserve the number of hollow nodes, we conclude that
the reduced stabilizer graphs for a stabilizer state are those that
have the least number of hollow nodes.

\section{Conclusion}
\label{sec:conclude}

Motivated by the relation between the graphs and the generator
matrices associated with graph states, we extend the graph
representation of states to encompass all stabilizer states.  These
\textit{stabilizer graphs\/} differ from the graphs employed by
\cite{hein} in that nodes can be either hollow or solid and can
possess both loops and signs.  The additional decorations identify
the local Clifford operations that relate the desired stabilizer
state to the graph state corresponding to the unadorned graph.
Imposing the restriction that the number of hollow nodes be minimal
yields a subset of the stabilizer graphs which we term reduced.
Reduced graphs follow naturally from the binary representation of the
stabilizer formalism, while generic stabilizer graphs are more
closely related to the quantum-circuit formalism.  For graph states,
reduced stabilizer graphs are identical to the standard
representation of these states in terms of graphs.

Using circuit identities, we derive a set of rules for transforming
stabilizer graphs under the application of various local Clifford
gates.  From this list, we abstract a similar set for transforming
reduced stabilizer graphs.  Considering these transformation rules,
particularly those for reduced stabilizer graphs, it becomes clear
that the mapping between stabilizer states and stabilizer graphs is
not one-to-one.  Rules for converting between equivalent graphs are
found, and we prove that the equivalence rules given are universal by
developing a constructive procedure for testing the equivalence of
any two stabilizer graphs.

\acknowledgments
This research was partly supported by Army Research
Office Contract No.~W911NF-04-1-0242 and National Science Foundation
Grant No.~PHY-0653596.

\appendix*

\section{Controlled-$Z$ gates\label{app:czgates}}
\begin{figure*}
\center
\includegraphics[width=18cm]{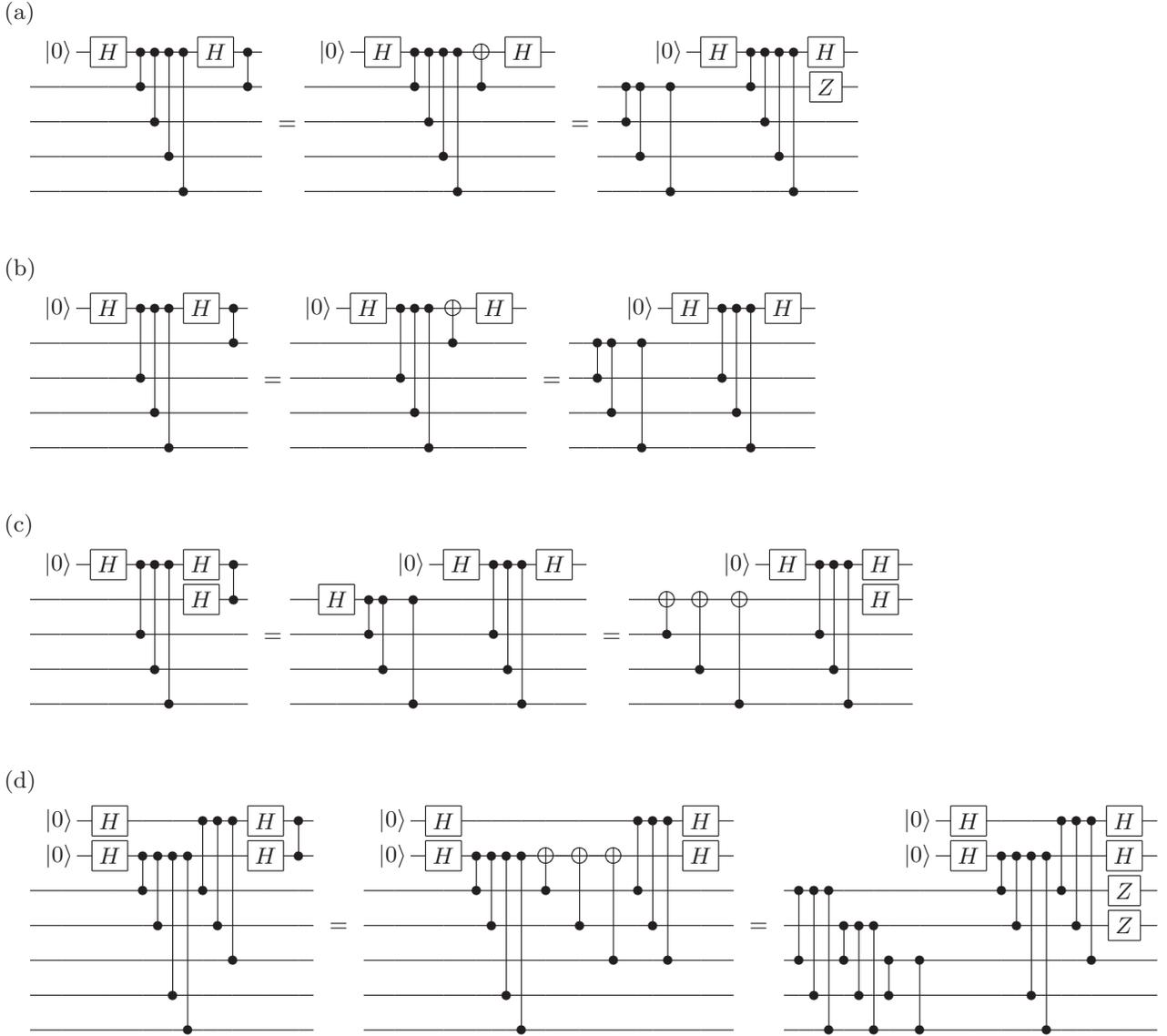}
\caption{Circuit identities for transforming a reduced stabilizer
graph under $\protect\CZ$ gates.  In~(a)--(c) the identity is
illustrated for the case of four neighbors connected to a qubit of
interest; equivalent expressions hold for other numbers of neighbors.
Identities~(a) and (b) give the rule~T(ix) for transforming a reduced
graph when a $\protect\CZ$ gate is applied to a hollow node and a
solid node. The effect of a $Z$ gate on the hollow qubit in the third
layer of~(a) or~(b) is to deposit an additional $Z$ gate on both the
hollow and solid qubits in the third layer of the final circuit. The
identity in~(c) relies on~(b); this identity is extended in~(d) to a
demonstration of the rule~T(x) for transforming a reduced graph when
a $\protect\CZ$ gate is applied to two hollow nodes.  If there is a
$Z$ gate on the lower hollow qubit in the third layer of~(c), the
result in the third layer of the final circuit is to put a $Z$ gate
on that qubit and on the neighbors of the other hollow qubit.
Translated to~(d), this means that a $Z$ gate in the third layer on
either hollow qubit leads in the third layer of the final circuit to
additional $Z$ gates on that qubit and on the neighbors of the other
hollow qubit.\label{fig:CZident}}
\end{figure*}

The transformation rules given in Sec.~\ref{sec:transformations}
suffice to describe the effect of any local Clifford operation on a
stabilizer state.  In order to complete the set of transformation
rules for the Clifford group, we include transformation rules for the
$\CZ$ gate here.  In the interest of brevity, we consider only
reduced-stabilizer-graph transformations.  Transformation rules for
general stabilizer graphs can be derived by first using the
equivalence rules in Sec.~\ref{subsec:circuitequiv} to convert to an
equivalent reduced graph and then applying the transformation rules
below.

\vspace{0pt}
\begin{list}{}{\itemindent=0pt\leftmargin=40pt\labelsep=5pt\labelwidth=40pt}

\item[T(viii).] Applying $\CZ$ between two solid nodes complements
the edge between them.

\vspace{0pt}

\item[T(ix).] Applying $\CZ$ between a hollow node and a solid node
complements the edges between the solid node and the neighbors of the
hollow node.\\
\rule{0em}{0em}\vspace{-.5em}\\
Flip the solid node's sign if the two nodes were initially connected
and the hollow node did not have a sign or if the two nodes were not
connected and the hollow did have a sign.

\vspace{0pt}

\item[T(x).] Applying $\CZ$ between two hollow nodes effects the
third step of local complementation along the (unoccupied) edge
between the two nodes.

Nodes that neighbor both decision nodes flip their signs.  If a
decision node initially had a sign, flip the signs of nodes connected
to the other decision node.
\end{list}

Transformation rule~T(viii) is trivial since the $\CZ$ gate simply
commutes into layer two of the reduced-graph-form circuit.  The
circuit identities needed to prove rules~T(ix) and~T(x) are given in
Fig.~\ref{fig:CZident}.

\end{document}